\documentclass[12pt]{article}
\usepackage{amssymb}
\usepackage{graphicx}
\usepackage[cp1251]{inputenc}
\usepackage{rotating}

\begin{document}
 \noindent {\footnotesize\it Astrophysical Bulletin, 2024, Vol. 79, No. 3, pp. 473--480}
 \newcommand{\dif}{\textrm{d}}

 \noindent
 \begin{tabular}{llllllllllllllllllllllllllllllllllllllllllllll}
 & & & & & & & & & & & & & & & & & & & & & & & & & & & & & & & & & & & & & &\\\hline\hline
 \end{tabular}

  \vskip 0.5cm
  \bigskip
 \bigskip
\centerline{\Large\bf Studying the Kinematics of the Stellar Association}
\centerline{\Large\bf TW Hya from Current Data}

 \bigskip
 \bigskip
  \centerline { 
   V. V. Bobylev\footnote [1]{bob-v-vzz@rambler.ru},  A. T. Bajkova}
 \bigskip
 \centerline{\small\it Pulkovo Astronomical Observatory, Russian Academy of Sciences, St. Petersburg, 196140 Russia}
 \bigskip
 \bigskip
{Abstract---The kinematics of the young stellar association near the Sun TW Hya is studied. Kinematic
estimates of the age of this association were obtained in two ways. The first method---analysis of stellar
trajectories integrated back in time---gave an estimate of the age $t=4.9\pm1.2$~Myr. The second was to
analyze the instantaneous velocities of stars and showed that there is a volume expansion of the stellar
system with an angular velocity coefficient $K_{xyz}=103\pm12$~km s$^{-1}$ kpc$^{-1}$. Based on this effect, the
time interval that passed from the beginning of the expansion of the TW Hya association to the present
moment was found, $t=9.5\pm1.1$~Myr. The following principal semi-axes of the residual velocity ellipsoid
are determined: $\sigma_{1,2,3}=(5.25,1.84,0.35)\pm(0.34,0.63,0.26)$~km s$^{-1}$.
 }

\bigskip
\section{Introduction}
There are several small associations and compact moving groups of very young (5--20 Myr) stars near the Sun. This is, for example, Tucana, $\beta$~Pic, $\eta$~Cha, $\epsilon$~Cha, TW~Hya. The emergence of all these objects appears to be closely related to the evolution of a larger and older complex, the Sco--Cen OB association.

The TW Hya association is one of the closest to us; its members are located at an average distance from the Sun of 60 pc. Most of its members are low-mass stars of spectral types K and M that have not reached the Main Sequence stage. The association study began in the 80--90s of the last century (de la Reza et al., 1989; Gregorio-Hetem et al., 1992; Zuckerman and Becklin, 1993). It currently includes more than 60 probable members (Luhman, 2023), a large percentage of which are binary or multiple systems. The age of the TW Hya association is estimated in (Luhman, 2023) as $10\pm2$~Myr. This value was found by averaging two estimates obtained from both kinematics analysis and photometry. In the first
case, the angular expansion coefficient was used, the value of which was found from the velocities of stars observed at present.

To estimate the age of the TWHya association, it is interesting to apply another kinematic method, which is based on the study of the orbits of stars integrated back into the past. This method has successfully estimated the kinematic ages of various open star clusters, moving groups and associations (for example, Yuan and Waxman, 1977; Fern\'andez et al., 2008; Couture et al., 2023). In particular, using this method for a still small number of members of the TW Hya association in the paper by de la Reza et al. (2006), an age estimate of $8.3\pm0.8$~Myr was obtained.

Due to its extreme proximity to the Sun, many kinematic and photometric parameters measured with very high accuracy are available to study this
association. For example, for almost all candidate stars, trigonometric parallaxes from the Gaia DR3 catalog (Vallenari et al., 2023) are known, obtained with relative errors of smaller than 3\%. The proper motions of the member stars of the TW Hya association are also measured in the Gaia DR3 catalog with very high accuracy.

But measurements of their radial velocities in the Gaia DR3 catalog are often made with larger errors compared to ground-based estimations. This is demonstrated in Bobylev and Bajkova (2024) using the moving group $\beta$ Pictoris as an example. Since many stars of the TW Hya association are part of multiple systems, reliable determination of the radial velocities of the center of mass requires long-term
observations, which so far can only be carried out using ground-based methods.

One of the important strengths of the Luhman (2023) study is that much work has been done to collect and analyze ground-based radial velocity measurements of candidate stars. As a result, the best measurements of stellar radial velocities to date are collected in one catalog.

The purpose of this work is to study the kinematics of the TW Hya association and kinematically
estimate its age. For this purpose, modern data on the probable members of the association are used
according to the list of Luhman (2023), where the parallaxes and proper motions of stars are taken from
the Gaia DR3 catalog, and the radial velocities are from literature sources. The method involves constructing the orbits of stars in the past at a given time interval and estimating the moment when the stellar group had a minimum spatial size.

\section{Method}
We use a rectangular coordinate system centered on the Sun, where the $x$ axis is directed towards the galactic center, the $y$ axis~--- towards galactic rotation and the $z$ axis~--- to the north pole of the Galaxy. Then $x=r\cos l\cos b,$ $y=r\sin l\cos b$ and $z=r\sin b,$ where $r=1/\pi$ is the heliocentric distance of the star in the kpc, which we calculate through the parallax of the star $\pi$ in mas (milliseconds of arc).

From observations, the radial velocity $V_r$ and two projected tangential velocity are known:
$V_l=4.74r\mu_l\cos b$ and $V_b=4.74r\mu_b,$  directed along
the galactic longitude $l$ and latitude $b,$ respectively, expressed in km s$^{-1}$. Here, the coefficient 4.74 is the
ratio of the number of kilometers in an astronomical
unit to the number of seconds in a tropical year. The components of proper motion, $\mu_l\cos b$ and $\mu_b$, are
expressed in milliseconds of arc year$^{-1}$.

Through the components $V_r, V_l, V_b,$ the velocities $U, V,W,$ are calculated, where the velocity $U$ is directed from the Sun to the center of the Galaxy, $V$ --- in the direction of the Galaxy rotation, and $W$ --- to the north galactic pole:
\begin{equation}
 \begin{array}{lll}
 U=V_r\cos l\cos b-V_l\sin l-V_b\cos l\sin b,\\
 V=V_r\sin l\cos b+V_l\cos l-V_b\sin l\sin b,\\
 W=V_r\sin b                +V_b\cos b.
 \label{UVW}
 \end{array}
 \end{equation}
Thus, the velocities $U,V,W$ are directed along the corresponding coordinate axes  $x,y,z$.

\subsection{Velocity ellipsoid}
To estimate the dispersions of residual velocities of stars we use the well-known method (Ogorodnikov,
1965), where six second-order moments are considered:
$a=\langle U^2\rangle-\langle U^2_\odot\rangle,$
 $b=\langle V^2\rangle-\langle V^2_\odot\rangle,$
 $c=\langle W^2\rangle-\langle W^2_\odot\rangle,$
 $f=\langle VW\rangle-\langle V_\odot W_\odot\rangle,$
 $e=\langle WU\rangle-\langle W_\odot U_\odot\rangle$ and
 $d=\langle UV\rangle-\langle U_\odot V_\odot\rangle,$
which are the coefficients of the surface equation
 \begin{equation}
 ax^2+by^2+cz^2+2fyz+2ezx+2dxy=1,
 \end{equation}
as well as components of the symmetric tensor of moments of residual velocities
 \begin{equation}
 \left(\matrix {
  a& d & e\cr
  d& b & f\cr
  e& f & c\cr }\right).
 \label{ff-5}
 \end{equation}
The following six equations are used to determine the values of this tensor:
\begin{equation}
 \begin{array}{lll}
 V^2_l= a\sin^2 l+b\cos^2 l\sin^2 l  -2d\sin l\cos l,
 \label{EQsigm-1}
 \end{array}
 \end{equation}
\begin{equation}
 \begin{array}{lll}
 V^2_b= a\sin^2 b\cos^2 l+b\sin^2 b\sin^2 l   +c\cos^2 b  \\
 -2f\cos b\sin b\sin l   -2e\cos b\sin b\cos l    +2d\sin l\cos l\sin^2 b,
 \label{EQsigm-2}
 \end{array}
 \end{equation}
\begin{equation}
 \begin{array}{lll}
 V_lV_b= a\sin l\cos l\sin b   +b\sin l\cos l\sin b\\
 +f\cos l\cos b-e\sin l\cos b   +d(\sin^2 l\sin b-\cos^2\sin b),
 \label{EQsigm-3}
 \end{array}
 \end{equation}
\begin{equation}
 \begin{array}{lll}
 V^2_r= a\cos^2 b\cos^2 l+b\cos^2 b\sin^2 l  +c\sin^2 b \\
 +2f\cos b\sin b\sin l   +2e\cos b\sin b\cos l +2d\sin l\cos l\cos^2 b,
 \label{EQsigm-4}
 \end{array}
 \end{equation}
\begin{equation}
 \begin{array}{lll}
 V_b V_r=-a\cos^2 l\cos b\sin b   -b\sin^2 l\sin b\cos b+c\sin b\cos b\\
 +f(\cos^2 b\sin l-\sin l\sin^2 b)   +e(\cos^2 b\cos l-\cos l\sin^2 b) \\
 -d(\cos l\sin l\sin b\cos b  +\sin l\cos l\cos b\sin b),
 \label{EQsigm-5}
 \end{array}
 \end{equation}
\begin{equation}
 \begin{array}{lll}
 V_l V_r=-a\cos b\cos l\sin l + b\cos b\cos l\sin l \\
    +f\sin b\cos l-e\sin b\sin l  + d(\cos b\cos^2 l-\cos b\sin^2 l),
 \label{EQsigm-6}
 \end{array}
 \end{equation}
which are solved by the least-squares method for six unknowns $a, b, c, f, e, d.$ Then the eigenvalues of the tensor (3) $\lambda_{1,2,3}$ are found from the solution of the secular equation
\begin{equation}
 \left|\matrix
 {
a-\lambda&          d&        e\cr
       d & b-\lambda &        f\cr
       e &          f&c-\lambda\cr
 }
 \right|=0.
 \label{ff-7}
 \end{equation}
The eigenvalues of this equation are equal to the reciprocal values of the squared semi-axes of the of
velocity moment ellipsoid and, at the same time, the squared semi-axes of the residual velocity ellipsoid:
 \begin{equation}
 \begin{array}{lll}
 \lambda_1=\sigma^2_1, \lambda_2=\sigma^2_2, \lambda_3=\sigma^2_3,\qquad
 \lambda_1>\lambda_2>\lambda_3.
 \end{array}
 \end{equation}
Directions of the main axes of the tensor (10)  $L_{1,2,3}$ and $B_{1,2,3}$ are found from the relations:
 \begin{equation}
 \tan L_{1,2,3}={{ef-(c-\lambda)d}\over {(b-\lambda)(c-\lambda)-f^2}},
  \end{equation}
 \begin{equation}
 \tan B_{1,2,3}={{(b-\lambda)e-df}\over{f^2-(b-\lambda)(c-\lambda)}}\cos L_{1,2,3}.
  \end{equation}
To calculate the residual velocities of stars, we use the values of the Sun's velocity relative to the local standard of rest
\begin{equation}
 (U_\odot,V_\odot,W_\odot)=(11.1,12.2,7.3)~\hbox {km s$^{-1}$},
 \end{equation}
obtained in the paper by Sch\"onrich et al. (2010). The spatial size of the TW Hya association is so small
that taking into account the differential rotation of the Galaxy is not required here. Moreover, it is not
necessary to take into account the influence of the spiral structure.

\subsection{Constructing the orbits of stars}
To construct the orbits of stars in the coordinate system rotating around the center of the Galaxy, we
use the epicyclic approximation (Lindblad, 1927):
\begin{equation}
 \renewcommand{\arraystretch}{1.8}
 \begin{array}{lll}\displaystyle
 x(t)= x_0+{U_0\over \displaystyle \kappa}\sin(\kappa t)   +{\displaystyle V_0\over \displaystyle 2B}(1-\cos(\kappa t)),  \\
 y(t)= y_0+2A \biggl(x_0+{\displaystyle V_0\over\displaystyle 2B}\biggr) t  -{\displaystyle \Omega_0\over \displaystyle B\kappa} V_0\sin(\kappa t)
      +{\displaystyle 2\Omega_0\over \displaystyle \kappa^2} U_0(1-\cos(\kappa t)),\\
 z(t)= {\displaystyle W_0\over \displaystyle \nu} \sin(\nu t)+z_0\cos(\nu t),
 \label{EQ-Epiciclic}
 \end{array}
 \end{equation}
where $t$ is the time in Myr (we proceed from the relationship 1~pc/1~Myr = 0.978~km s$^{-1}$), $A$ and $B$ are the
Oort constants; $\kappa=\sqrt{-4\Omega_0 B}$ is the epicyclic frequency; $\Omega_0$ is the angular velocity of the galactic
rotation of the local rest standard, $\Omega_0=A-B$; $\nu=\sqrt{4\pi G \rho_0}$ is the vertical vibration frequency,
where $G$ is the gravitational constant, $\rho_0$ is the stellar density in the solar neighborhood.

Parameters $x_0,y_0,z_0$ and $U_0,V_0,W_0$ in the systemof equations (15) denote the current positions and
velocities of the stars, respectively. The rise of the Sun above the galactic plane $h_\odot$ is taken to be 16 pc according to the paper by Bobylev and Bajkova (2016). We calculate the velocities $U,V,W$ relative to the local rest standard using the values (14). We accepted $\rho_0=0.1~M_\odot/$pc$^3$  (Holmberg and Flynn, 2004), what gives $\nu=74$~km s$^{-1}$ kpc$^{-1}$. We use the following
values of the Oort constants: $A=16.9$~km s$^{-1}$ kpc$^{-1}$ and $B=-13.5$~km s$^{-1}$ kpc$^{-1}$ which are close to
modern estimates. A review of such estimates can be found, for example, in the paper by Krisanova et al. (2020).

\section{Data}
The basis of our sample is formed by stars of the TW Hya association from the Luhman list (2023).
This list contains 67 probable members of the association, which are members of 55 binary or multiple
systems. A number of stars in the list do not have radial velocity measurements. Ultimately, we selected 53 stars with measured parallaxes, proper motions, and radial velocities for kinematic analysis.

Luhman (2023) included in his catalogue both the trigonometric parallaxes from the Gaia DR3 catalogue and the distances to stars from Bailer-Jones et al. (2021). The latter were precisely what Luhman used in his calculations. It should be noted
that the trigonometric parallaxes from the Gaia DR3
catalogue for all analyzed stars were measured in the indicated paper with relative errors of smaller than 3\%. With such errors in the distances to stars calculated directly through the parallax $r = 1/\pi,$ it is
possible to ignore any model-dependent additional corrections (Lutz and Kelker, 1973). Therefore, in
this paper we use distances to stars calculated directly using trigonometric parallaxes from the Gaia DR3 catalogue.

\begin{figure}[t]
{ \begin{center}
  \includegraphics[width=0.9\textwidth]{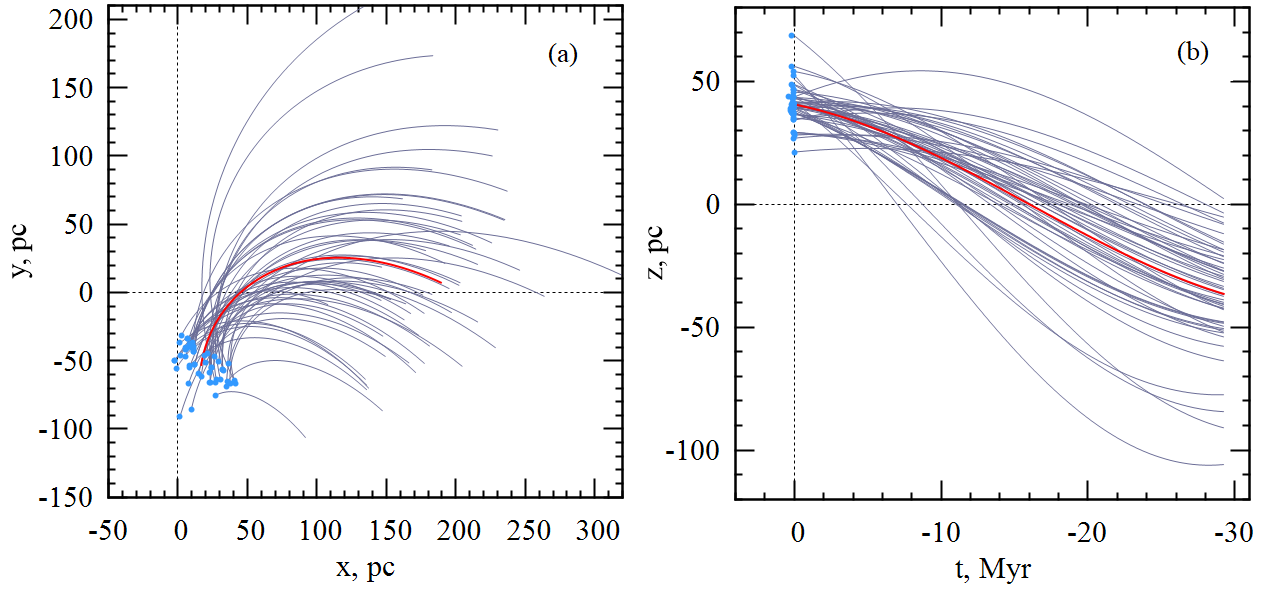}
  \caption{Distribution of 53 stars of the TW Hya association projected onto the galactic $xy$ plane and their trajectories (a), the
vertical distribution and their trajectories (b), the trajectories traced back into the past over an interval of 30~Myr, the trajectory
of the kinematic center of this stellar group shown in red.}
 \label{f1-DR3}
\end{center}}
\end{figure}
\begin{figure}[t]
{ \begin{center}
   \includegraphics[width=0.75\textwidth]{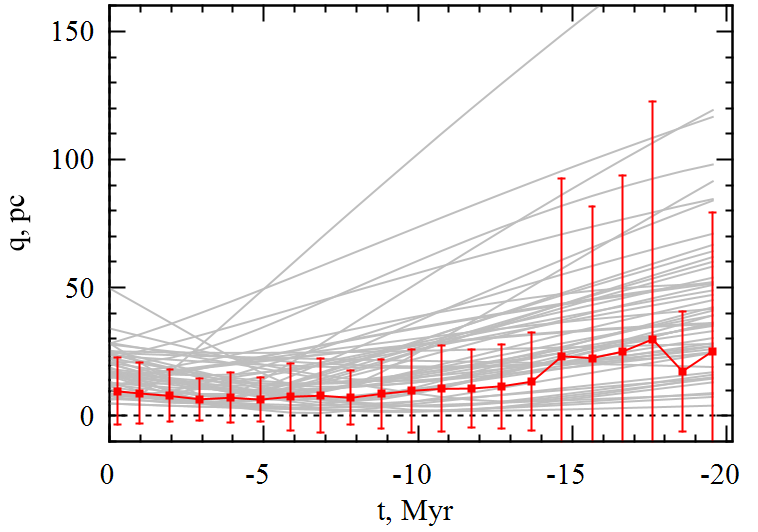}
  \caption{Deviations from the trajectory of the kinematic center (parameter $q$) on an integration interval of 20~Myr
for 53 stars of the TW Hya association; average values and corresponding dispersions are shown in red.}
 \label{f2-q}
\end{center}}
\end{figure}
\begin{figure}[t]
{ \begin{center}
   \includegraphics[width=0.98\textwidth]{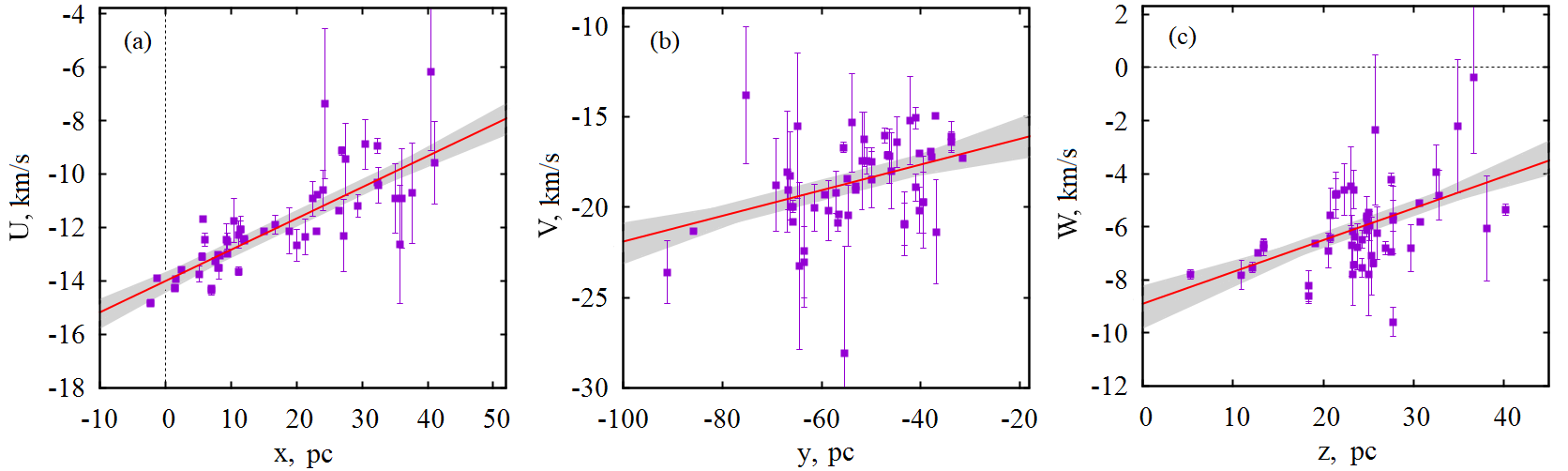}
  \caption{Velocities $U$ depending on the $x$ coordinate (a), velocities $V$ depending on the $y$ coordinate (b), and velocities $W$
depending on the $z$ coordinate (c) for 53 stars of the TW Hya association.}
 \label{f6-Keff}
\end{center}}
\end{figure}

\section{Results and discussion}
Figure 1 shows the current positions (blue circles) and trajectories from the past of 53 stars in the TW Hya association. The integration of the orbits of the stars was performed over an interval of 30 Myr back into the past.

The characteristics of the kinematic center of this group were also obtained. The trajectory of the kinematic center is defined as follows. We calculate the average values of the positions and velocities of the stellar group:
${\overline x}_0,{\overline y}_0,{\overline z}_0$ and ${\overline U}_0,{\overline V}_0,{\overline W}_0$. Thus,
from the data for 53 stars of the TWHya association:
${\overline U}_0=-12.1$~km s$^{-1}$,
${\overline V}_0=-18.8$~km s$^{-1}$ and
${\overline W}_0=-6.0$~km s$^{-1}$ and also
${\overline x}_0=16.7$~pc,
${\overline y}_0=-53.0$~pc and
${\overline z}_0=24.7$~pc,
which are almost identical to similar values from Luhman (2023) calculated using a slightly different sample of stars.

Using these values, the trajectory of the kinematic center is constructed. Based on the differences in coordinates (between the star and the kinematic center) $\Delta x,\Delta y,\Delta z$ at each moment of integration for each star we determine the value of the parameter $q$ of the
following form:
\begin{equation}
 q=\sqrt{\Delta x^2+\Delta y^2+\Delta z^2},
 \label{qq}
 \end{equation}
which characterizes the deviation of the star from the trajectory of the kinematic center, which is also
shown in Fig. 1. Note that the trajectories of the
stars are calculated taking into account the elevation of the Sun above the galactic plane. Thus, in all our figures the $z$ coordinate reflects the position of the stars relative to the plane of the Galaxy.

For each star, the dependence of the parameter $q$ on time is constructed over an integration interval of 20 Myr (Fig. 2). A shorter integration interval is taken compared to the previous step in order to be able to track the position of the minimum of the average values in more detail. It is clearly visible that

i)~the spatial size of the stellar group was significantly larger 15--20 Myr ago compared to the present;

ii)~the red line in Fig. 2 has a minimum in the region of 5 million years;

iii)~there is a tendency for the stellar group to expand.

From the analysis of the averaged data, the following age estimation of the association TW Hya was obtained:
 \begin{equation}
 t=4.9\pm1.2~\hbox {Myr}.
 \label{t-DR3}
 \end{equation}
The error of the moment t was found as a result of statistical modeling using the Monte-Carlo method.
It was assumed that the orbits of the stars were constructed with relative errors of 10\% distributed according to the normal law.

Figure 3 shows the dependences of the velocities $U, V,$ and $W$ on the corresponding coordinates $x, y,$
and $z,$ as well as the gradients found from these data using the least squares method:
 $$\partial U/\partial x=117\pm11~\hbox {km s$^{-1}$ kpc$^{-1}$}, $$
 $$\partial V/\partial y=  71\pm23~\hbox {km s$^{-1}$ kpc$^{-1}$}, $$
 $$\partial W/\partial z=120\pm26~\hbox {km s$^{-1}$ kpc$^{-1}$} $$
with indication of the boundaries of confidence intervals corresponding to level 1$\sigma.$ Note that the calculation of gradient values was performed in several iterations with the exclusion of large discrepancies according to the 3$\sigma$ criterion. In each of the three cases mentioned above, no more than one star was rejected based on this criterion.

In fact, based on the data of the three panels presented in the figure, three gradients were
found, which in the linear Ogorodnikov-Milne model (Ogorodnikov, 1965; Bobylev and Bajkova, 2023)
are diagonal members of the deformation matrix and describe the effects of the expansion of the stellar
system. As a result, we can evaluate the effect of volume expansion of the association TW Hya,
 $K_{xyz}=(\partial U/\partial x+\partial V/\partial y+\partial W/\partial z)/3$:
 \begin{equation}
K_{xyz}=(\partial U/\partial x+\partial V/\partial y+\partial W/\partial z)/3=103\pm12~\hbox {km s$^{-1}$ kpc$^{-1}$}
 \label{t-xyz}
 \end{equation}
and find the time interval that has passed from the beginning of the expansion of the star system to the
present moment, $t=977.5/K_{xyz}$:
 \begin{equation}
 t=9.5\pm1.1~\hbox {Myr.}
 \label{t-Kxyz}
 \end{equation}
The spatial size of the TWHya association does not exceed 60--70 pc (Fig. 1a). It is interesting to estimate what kind of linear expansion velocity of
the association at its outer boundary $V_{\rm exp}$ we can talk about. Let us take the radius of the association
$R_{\rm TWA}=30$~pc, then $V_{exp}=K_{xyz} R_{\rm TWA}=3$~km s$^{-1}$. This value is in agreement with known estimates of the expansion rates of various stellar associations, 3--6~km s$^{-1}$ (Mel'nik and Dambis, 2017, 2018; Wright, 2020).

As a rule, from the analysis of OB associations, either a flat effect (expansion in a plane) or an expansion effect along one direction is known. Let us
note the paper by Bobylev and Bajkova (2023), in which the coefficient of volume expansion was found
$K_{xyz}=43.2\pm3.4$~km s$^{-1}$ kpc$^{-1}$ in an analysis of a large sample of stars---members of the Sco--Cen association.

In the paper by Luhman (2023), the age of the TW Hya association was estimated in two ways. First, a kinematic estimation was obtained from the expansion rate $t=9.6_{-0.8}^{+0.9}$~Myr. And it is made on the basis of the coefficient of volume expansion $K_{xyz}$ with the value $102\pm9$~km s$^{-1}$ kpc$^{-1}$. The difference from our result (18) is explained by the different number of stars used for the calculation. Second, from the position of the combined sequence of low-mass stars on the Hertzsprung--Russell diagram, it was found that $t=11.4_{-1.2}^{+1.3}$~Myr. And based on these two results, in the paper by Luhman (2023) the final estimation $t=10\pm2$~Myr was adopted.

Using 53 stars of the TW Hya association, the
following parameters of the residual velocity ellipsoid
were found in this paper:
\begin{equation}
 \begin{array}{lll}
  \sigma_1= 5.25\pm0.34~\hbox{km s$^{-1}$}, \\
  \sigma_2= 1.84\pm0.63~\hbox{km s$^{-1}$}, \\
  \sigma_3= 0.35\pm0.26~\hbox{km s$^{-1}$}
 \label{rez-6}
 \end{array}
 \end{equation}
and the orientation parameters of this ellipsoid:
 \begin{equation}
  \matrix {
  L_1=~~2\pm6^\circ, & B_1=12\pm6^\circ, \cr
  L_2=~92\pm1^\circ, & B_2=~0\pm2^\circ, \cr
  L_3=181\pm1^\circ, & B_3=78\pm1^\circ. \cr
   }
 \label{rez-66}
 \end{equation}
In the considered neighborhood of the Sun, the errors in the velocities $V_l$ and$V_b$ have significantly smaller
values compared to the errors $V_r$ (the effect is clearly visible, for example, in Fig. 1 of the paper by Bobylev et al., 2021). Therefore, parameters (20) were found using only the proper motions of the stars (using three equations (4)--(6)).

For comparison, from the solution of the complete system of equations (4)--(9), the following velocities were obtained:
  $$
  \sigma_{1,2,3}=(7.86,3.06,0.86)\pm(0.66,0.99,0.82)~\hbox{km s$^{-1}$}.
  $$
The orientation of the axes is as follows:
 $(L,B)_1=(72^\circ,-18^\circ),$
 $(L,B)_2=(167^\circ,-14^\circ)$ and
 $(L,B)_3=(114\pm4^\circ,67\pm3^\circ)$
 (if the errors in determining the direction are large, then we do not show them). Note that the parameters of this ellipsoid are in good agreement with the data in Fig. 1, where the initial velocities were found for all three components $V_l, V_b$ and $V_r$.

The errors of each axis of the residual velocity ellipsoid are estimated independently according to relations (12) and (13). We can see that the orientation of the first axis of the ellipsoid (21) is poorly determined.
This is caused by the narrow range of values of the
galactic longitudes, but one can focus on the value of the third axis ($B_3$) which is determined with small errors.

The values of parameters (20) are close to the corresponding values of the velocities of stars of the
Sco--Cen association. For example, Bobylev and Bajkova (2020) showed that the residual velocity
ellipsoid of the Sco--Cen association stars has semi-major axes:
$$\sigma_{1,2,3}=(7.72,1.87,1.74)\pm(0.56,0.37,0.22)~\hbox{km s$^{-1}$},$$
and it is located at an angle of $22\pm2^\circ$ ($B_3=78\pm2^\circ$)
to the vertical. It is also well known that the third axis of the Gould Belt has a tilt close to that found (Bobylev, 2014; Bobylev and Bajkova, 2020). By the way, the famous Gould Belt expansion effect is established in the paper by (Bobylev, 2014; Bobylev and Bajkova, 2020) based on the flat expansion coefficient $K_{xy}$ (expansion in the $xy$ plane).

In the paper by Foster et al. (2015), the authors used high-accuracy radial velocity measurements of stars to study the evolutionary status of the
young (1--2 Myr) open star cluster NGC~1333. These authors obtained the dispersion of stellar velocities
$\sigma_r=0.92\pm0.12$~km s$^{-1}$ taking into account the correction for binarity and came to the conclusion that the cluster is close to an equilibrium state.

We note the paper by Wei et al. (2024), where for the core of a young open star cluster in the Orion Nebula (the age is about 2 Myr) an estimation of the dispersion of internal velocities was obtained: $\sigma_{1D_{3D}}=2.26\pm0.08$~km s$^{-1}$ which, according to
these authors, indicates a deviation of the cluster from the equilibrium state (for the equilibrium state a smaller value of $\sigma_{1D_{3D}}$ is required). Here, the dispersion was found as the root-mean-square value in three directions ($\alpha,\delta,r$).

The similar value calculated for the association TW Hya is $\sigma=3.22\pm0.44$~km s$^{-1}$ which indicates the extreme youth of the TW Hya association. Moreover, due to the significant percentage of multiple systems in the sample, the value $\sigma$ may be overestimated. However, the presence of a reliably established expansion directly indicates that the TW Hya association is far from the equilibrium state.

\section{Conclusions}
A sample of probable members of the TW Hya association was formed according to the list in Luhman
(2023). It included 53 stars with trigonometric parallaxes, proper motions from the Gaia DR3 catalogue,
as well as radial velocities taken from the literature sources.

To estimate the kinematic age of the TW Hya association, the orbits of stars in the past were constructed and the moment, when the stellar group had a minimum spatial size, was determined. As a result, the age of the association was estimated in two ways. In both cases, the estimates are kinematic.

The study of the trajectories of stars constructed for the past has provided the estimate $t=4.9\pm1.2$~Myr
(result (17)). The analysis of the instantaneous velocities of stars allows us to speak about the volume
expansion of this stellar system. Based on this effect, the time interval that has passed from the beginning
of the expansion of the   association to the present moment was found, $t=9.5 \pm1.1$~Myr (result (19)).

It is interesting that the found coefficient of the volume expansion of the association TW Hya differs significantly from zero, $K_{xyz}=103\pm12$~km s$^{-1}$ kpc$^{-1}$. Of course, this was possible due to the highmeasurement accuracy of the positions and velocities of the stars. Typically, from the analysis of OB associations,either a flat effect (the expansion in a plane) or an expansion effect along a single direction is known. This is because the gradient $\partial W/\partial z$ is usually very poorly defined.

The following values of the main semi-axes of the residual velocity ellipsoid were found for the stars of the TW Hya association:
$\sigma_{1,2,3}=(5.25,1.84,0.35)\pm(0.34,0.63,0.26)$~) ~km s$^{-1}$ . Such values of velocity dispersions are characteristic
of internal motions in molecular cloud complexes, as well as very young stellar associations and clusters. The third axis of the found velocity ellipsoid is deflected by $12\pm1^\circ$ of the vertical. It is note worthy that such a deflection value is typical of larger formations known in the solar neighborhood, the Sco--Cen association and the Gould belt.

 \subsubsection*{Acknowledgments}
The authors are grateful to the reviewer for useful comments that contributed to improving the paper.

 \subsubsection*{REFERENCES}
 \small

\quad~1. C. A. L. Bailer-Jones, J. Rybizki, M. Fouesneau, et al., Astron. J. 161 (3), id. 147 (2021). DOI: 10.3847/1538-3881/abd806

2. V. V. Bobylev, Astrophysics 57 (4), 583 (2014). DOI: 10.1007/s10511-014-9360-7

3. V. V. Bobylev and A. T. Bajkova, Astronomy Letters 42 (1), 1 (2016). DOI: 10. 1134/S1063773716010023

4. V. V. Bobylev and A. T. Bajkova, Astrophysical Bulletin 75 (3), 267 (2020). DOI: 10.1134/S1990341320030025

5. V. V. Bobylev and A. T. Bajkova, Astronomy Letters 49 (7), 410 (2023). DOI: 10.1134/S1063773723070010

6. V.V. Bobylev and A.T. Bajkova,  Astronomy Letters 50 (4), 239 (2024). DOI: 10.1134/S1063773724700117

7. V. V. Bobylev, A. T. Bajkova, A. S. Rastorguev, and M. V. Zabolotskikh, Monthly Notices
Royal Astron. Soc. 502 (3), 4377 (2021). DOI: 10.1093/mnras/stab074

8. V. V. Bobylev and A. T. Baykova, Astronomy Reports 64 (4), 326 (2020). DOI: 10.1134/S1063772920040022

9. D. Couture, J. Gagn\'e, and R. Doyon, Astrophys. J. 946 (1), id. 6 (2023). DOI: 10.3847/1538-4357/acb4eb

10. R. de la Reza, E. Jilinski, and V. G. Ortega, Astron. J. 131 (5), 2609 (2006). DOI: 10.1086/501525

11. R. de la Reza, C. A. O. Torres, G. Quast, et al., Astrophys. J. 343, L61 (1989). DOI: 10.1086/185511

12. D. Fern\'andez, F. Figueras, and J. Torra, Astron. and Astrophys. 480 (3), 735 (2008). DOI: 10.1051/0004-6361:20077720

13. J. B. Foster, M. Cottaar, K. R. Covey, et al., Astrophys. J. 799 (2), id. 136 (2015). DOI: 10.1088/0004-637X/799/2/136

14. J. Gregorio-Hetem, J. R. D. Lepine, G. R. Quast, et al., Astron. J. 103, 549 (1992). DOI: 10.1086/116082

15. J.Holmberg and C. Flynn, Monthly Notices Royal Astron. Soc. 352 (2), 440 (2004). DOI: 10.1111/j.1365-2966.2004.07931.x

16. O. I. Krisanova, V. V. Bobylev, and A. T. Bajkova, Astronomy Letters 46 (6), 370 (2020). DOI: 10.1134/S1063773720060067

17. B. Lindblad, Arkiv for Matematik, Astronomi och Fysik 20A, No. A.17 (1927).

18. K. L. Luhman, Astron. J. 165 (6), id. 269 (2023). DOI: 10.3847/1538-3881/accf19

19. T. E. Lutz and D. H. Kelker, Publ. Astron. Soc. Pacific 85 (507), 573 (1973). DOI: 10.1086/129506

20. A. M. Mel'nik and A. K. Dambis, Monthly Notices Royal Astron. Soc. 472 (4), 3887 (2017).
DOI: 10.1093/mnras/stx2225

21. A. M. Mel'nik and A. K. Dambis, Astronomy Reports 62 (12), 998 (2018). DOI: 10.1134/S1063772918120089

22. K. F. Ogorodnikov, Dynamics of stellar systems (Pergamon Press, New York, 1965).

23. R. Sch\"onrich, J. Binney, and W. Dehnen, Monthly Notices Royal Astron. Soc. 403 (4), 1829 (2010).
DOI: 10.1111/j.1365-2966.2010.16253.x

24. L. Wei, C. A. Theissen, Q. M. Konopacky, et al., Astrophys. J. 962 (2), id. 174 (2024). DOI: 10.3847/1538-4357/ad1401

25. A. Vallenari et al. (Gaia Collab.), Astron. and Astrophys. 674, id. A1 (2023). DOI: 10.1051/0004-6361/202243940

26. N. J. Wright, New Astron. Rev. 90,  id. 101549 (2020). DOI: 10.1016/j.newar.2020.101549

27. C. Yuan and A. M. Waxman, Astron. and Astrophys. 58 (1--2), 65 (1977).

28. B. Zuckerman and E. E. Becklin, Astrophys. J. 406, L25 (1993). DOI: 10.1086/186778

\end{document}